\documentclass[preprint,aps,nofootinbib,11pt]{revtex4-1}
\usepackage{geometry}                
\usepackage{amsmath,amssymb,amsfonts,dcolumn,color,graphicx,graphics,latexsym,placeins,epsfig,tikz}
\usetikzlibrary{arrows,shapes}
\usepackage{subfigure,rotating,bm,mathrsfs}
\usepackage[title]{appendix}

\newcommand{\be}{\begin{equation}}
\newcommand{\ee}{\end{equation}}
\newcommand{\ba}{\begin{eqnarray}}
\newcommand{\ea}{\end{eqnarray}}

\usepackage[pagebackref=false, colorlinks=true]{hyperref}
\definecolor{redish}{rgb}{0.7,0.2,0.0}  
\definecolor{bluish}{rgb}{0.2,0.5,0.8}

\hypersetup{linkcolor=redish,          
                  citecolor=blue,        
                  filecolor=magenta,      
                  urlcolor=blue}          

\begin{document}

\author{Kuntal Pal}
\email{kuntal@iitk.ac.in}
\author{Kunal Pal}
\email{kunalpal@iitk.ac.in}
\author{Ankit Gill}
\email{ankitgill20@iitk.ac.in}
\author{Tapobrata Sarkar}
\email{tapo@iitk.ac.in}
\affiliation{Department of Physics, Indian Institute of Technology, Kanpur 208016, India}

\title{\Large Evolution of circuit complexity in a harmonic chain under multiple quenches}

\begin{abstract}
We study Nielsen's circuit complexity in a periodic harmonic oscillator chain, under single and
multiple quenches. In a multiple quench scenario, it is shown that
the complexity shows remarkably different behaviour compared to the other information theoretic
measures, such as the entanglement entropy. In particular, after two successive quenches, when 
the frequency returns to its initial value, there is a lower limit of complexity, 
which cannot be made to approach zero. Further, we show that by applying a large number of successive 
quenches, the complexity of the time evolved state can be increased to a high value, which is not 
possible by applying a single quench. This model also exhibits the interesting phenomenon of 
crossover of complexities between two successive quenches performed at different times. 
\end{abstract}
\maketitle

\section{Introduction}

Nonequilibrium dynamics of quantum systems after a quench of the system
parameters is paradigmatic of the current thrust in research on quantum information theory,
and is expected to be of fundamental importance in technological applications of 
quantum theory. While there is a large number of ways to take a system out of equilibrium and then
study its evolution, quantum quenches are ubiquitous in this regard and have played a
vital role in our understanding of nonequilibrium dynamics of closed quantum systems \cite{Polkovnikov}. 
Here, the final fate of the system depends on its integrable or chaotic nature, 
and various physical measures have been used to characterise such a final state \cite{DAlessio}. 
To this end, the reduced density matrix, entanglement measures calculated from the 
reduced density matrix, such as entanglement entropy (EE), 
entanglement negativity etc., as well as information theoretic quantities like mutual 
information have been widely used to study thermodynamic equilibrium after a quench \cite{Peschel1}-\cite{Alba}. 

On the other hand, it is essential to understand how information spreads across all 
the degrees of freedom of a many body system after the dynamical equilibrium of the 
system has been disturbed, and this seems to strongly depend on the integrable
properties of the Hamiltonian. The notion of information scrambling, which is roughly the fact that 
any local perturbation will not be strictly local during the course of time
evolution and can only be measured  globally, is generally ascribed to chaotic systems. 
The canonical prescription to see the effect of scrambling, namely the out of time ordered
correlator (OTOC) has been studied in a variety of systems mainly by using numerical methods
\cite{Larkin}-\cite{Shukla:2022}. Even though the general consensus is that the saturation of OTOC 
at late times indicate scrambling of information in chaotic systems, 
there are also results in the literature that point to the contrary. Among the
significant ones, it was shown in \cite{Lin}, by means of computing OTOC (in an integrable Ising chain), and 
in \cite{Alba} by computing the algebraic decay of mutual information (in spin $\frac{1}{2}$ chain), 
that integrable models also can show information scrambling. The authors of \cite{Xu}, 
on the other hand proved that OTOC of integrable systems that have unstable fixed
points in phase space, can also show scrambling, therefore the notion of scrambling 
may not always be synonymous of chaos. 
To further support this point, one has to only point out that even a trivially integrable system 
like a collection of interacting simple harmonic oscillators gives a saturation of OTOC
under a multiple quench protocol \cite{GGS2}. 

In the view of above discussion, in this paper our focus will be to study 
a relatively new information theoretic quantity, the circuit complexity (CC), 
in integrable systems that shows scrambling under multiple quenches. 
The notion of CC, that measures the number of gates 
it is required to create a desired target state starting from a given reference state, has
recently gained a widespread attention in literature, specially because it can offer a probe to the
black hole interiors. On the quantum mechanical side, the most popular method of quantifying
the CC is due to Nielsen and collaborators 
\cite{Nielsen1}-\cite{Nielsen3}, who developed a geometrical formulation on the space
unitary operators to calculate the  circuit complexity.
The Nielsen complexity (NC), that measures the geodesic distance 
between two points on the unitary Riemannian manifold for a particular cost functional, was used in 
\cite{Myers1}-\cite{HM} for possible extensions in quantum field theories. 
The time evolution of NC after a quantum quench in integrable systems as studied in 
\cite{Alves}, shows linear growth followed by saturation, which generically 
captures the underlying physics of the system. 
Quench in the context of complexity have been also studied in \cite{CCDHJ}, \cite{Liu}.
On the other hand, with a slightly different motivation, the complexity between a 
reference state and the target state obtained by forward time evolution with a
particular Hamiltonian followed by backward evolution with slightly different Hamiltonian 
shows markedly different behaviour, depending on whether the classical Hamiltonian is 
chaotic or not \cite{Ali}. For a classically exactly solvable model such as a collection of 
harmonic oscillators, the complexity
shows usual oscillatory behaviour, but for a classically chaotic system like the
inverted harmonic oscillator, the complexity shows linear growth after some time, 
there by making CC a good measure of the dynamics of the quantum chaotic system 
\cite{Ali}, \cite{Qu}. Here we study the CC of interacting harmonic chain, 
which  in terms of dynamics lies somewhat in between 
the above mentioned two examples of simple integrable systems and a fully chaotic system, 
as this integrable  model we consider shows
scrambling of information between different degrees of freedom \cite{GGS2}. 
The evolution of complexity is studied both under a single and the above mentioned multiple quench protocols.
Our motivation here is two fold. 
Firstly, we want to study how a complicated measure like CC, which depends on the reference state, the
target state, as well as on the cost functional chosen,  
behaves in multiple quench scenario and its comparison with the EE. 
To be specific, it is interesting to ask if the employment of more than one quenches 
can change the saturation time scales of EE and CC, as it has already been observed that 
the EE can saturate slowly compared to CC in a single quench scenario \cite{Ali2}.  
Our second aim is to see how CC of the full system behaves after the local information 
has already been distributed across the whole system, which in principle is encoded in the wave function. 
As the final state of this system points to a relaxation behaviour from the analysis of 
momentum distribution and EE, it is useful to see if the analysis of CC also gives the same conclusion.
Our method is fully analytical and can be generalised for arbitrary time-dependent frequency
and interaction strength profile of the harmonic chain.

The main results that we obtain here are as follows. For a single quench, we find that 
as expected, complexity shows revivals, whose time periods depend on the post-quench
frequency of the oscillators. 
Indeed, dynamically generated time scales  are responsible for these quasi revivals. 
Further, when such a quench is critical, the complexity grows at large times because of the 
logarithmic divergence of the zero mode contribution at this limit. 
Finally, when the zero mode contribution is  subtracted from the complexity, it does not show any
decoherence in time, which is contrary to what is observed in the same system, when the 
covariance matrix is used to calculate the complexity \cite{dGT2,dGT1}. 

For multiple quenches, we find that, unlike the entanglement entropy, the complexity does not 
attain a steady state value even after  multiple quenches, but rather continues to oscillates
even at large times. Further, after two successive quenches, when the reference state 
frequency goes back to initial value, the complexity has a finite value which is 
higher than the complexity just after the next quench. There is a limit on how close a time 
evolved state can be closer to the reference state before the quench, and complexity 
can not be made to approach close to zero by increasing the frequency, which controls its magnitude.
In this regard, we find a novel phenomenon which is not possible to observe with a single quench, namely
a quench  performed at an earlier  time (i.e., a quench with lower quench number $i$) `pushes away'
the time evolved target state far away on the space of unitaries compared to a quench performed
at a later time. Hence there is a crossover of  complexities of successive quenches just after the $(i+1)$th
quench. Also, by using a sufficiently large number of quenches, a target state with large complexity can be 
prepared, which is in contrast with the single quench where the peaks of the complexity maxima are 
fixed by the quench frequency. These results are elaborated upon in the following sections. 

\section{Complexity evolution in a harmonic chain under a single quench}

We  consider a chain containing $N$ mutually interacting harmonic oscillators (``harmonic chain'') whose frequency
depends on time. Furthermore, the interaction strength is also assumed to be dependent on time.
We will study the time evolution of Nielsen's circuit complexity of such a system, when the system parameters, namely 
the frequency and the interaction strengths are quenched to different values under some definite protocols, which
we will define in sequel. Complexity evolution
of such a systems under a global quench has been studied in \cite{dGT1} using the covariance matrix 
method \cite{dGT2}. Here we shall employ the wavefunction method of calculating the circuit complexity, and use 
the solution of the Ermakov-Milne-Pinney (EMP) equation to write down the corresponding complexity. We will also 
discuss the differences in the resulting complexity computed using these two methods.

\subsection{Ground state of the harmonic chain}
The Hamiltonian under consideration is that of $N$ mutually interacting harmonic oscillators 
whose frequency and interaction strength are explicit functions of time, and is given as
\begin{equation}
H(x_j,p_j,t)=\frac{1}{2}\bigg[\sum_{j=1}^{N}\Big(p_j^2+\omega(t)^2x_j^2\Big)+k(t)\Big(x_j-x_{j+1}\Big)^2\bigg]
=\frac{1}{2}\bigg[\sum_{j=1}^{N}\Big(p_j^2+X^T\cdot K(t) \cdot X\Big)\bigg]~.
\end{equation}
Here $X=(x_1,x_2,...x_N)^T$ is a column matrix for the collective position of each oscillator, and $K$ is 
a real symmetric matrix whose  eigenvalues are denoted as $\lambda_j$. It is assumed that
periodic boundary conditions are imposed on the chain.

By using an orthogonal transformation $U$, we can diagonalise the above Hamiltonian so that
each of the $N$ oscillators become decoupled.
Denoting $Y=(y_1,y_2,...y_N)^T=UX$	as the new coordinates and
$K^D=UKU^T$ as the diagonal form of the matrix $K$, the transformed Hamiltonian can be written as 
\begin{equation}\label{Nparticlediagonal}
H^D(y_j,P_j,t)=\frac{1}{2}\bigg[\sum_{j=1}^{N}\Big(P_j^2+ K^D_{jj}(t) y_j^2\Big)\bigg]~.
\end{equation}
We are interested in the ground state of the system.
Since the Hamiltonian is an explicit function of time, hence to find the ground state of the system we can use
the Lewis-Risenfeld theory of time-dependent invariant operators.
In diagonal coordinates $y_j$, the ground state wavefunction corresponding to
the Hamiltonian of Eq. (\ref{Nparticlediagonal}) can be written as \cite{GGS1}
\begin{equation}\label{groundstate}
\Psi_0(y_j,t)=\bigg(\prod_{j=1}^{N}\frac{1}{b_j^2(t)}\text{det}\frac{\sqrt{K^D(t)}}{\pi}\bigg)^{1/4}
\exp\bigg[\sum_{j=1}^{N}\bigg(-\frac{1}{2}\Omega_j(t) y_j^2+\alpha_j(t)\bigg)\bigg]~,~
\Omega_j (t)=\frac{\sqrt{K^{D}_{jj}(t=0)}}{b_{j}^{2}(t)}-i\frac{\dot{b}_j(t)}{b_j(t)}~.
\end{equation}
Here $\alpha_j(t)$ is a time-dependent phase factor whose form will not be important to us, and an overdot indicates
a derivative with respect to time.
The auxiliary functions $b_j(t)$ satisfies the following  EMP equation 
\begin{equation}\label{auxiliary_eq}
\ddot{b}_j+\lambda_j(t)b_j-\frac{\lambda_j(0)}{b_j^3(t)}=0~,
\end{equation}
where $\lambda_j(t)=\omega^2(t)+2k(t)\big[1-\cos(\frac{2\pi j}{N})\big]$, with $j=1,2,\cdots, N$, are the time-dependent 
eigenvalues of the matrix $K$, and $\lambda_j(0)=\lambda_j(t=0)$.
To determine the unknown constants 
appearing in the general solution of this equation, we need to impose conditions on it at some initial time, say $t=t_0$.
Here, we shall impose the  following  conditions on $b(t)$ at the start of the quench $t=0$ : 
\begin{equation}\label{initial}
b(t=0)=1~ \quad \text{and}\quad \dot{b}(t=0)=0~,
\end{equation}
so that the wave function at $t=0$ matches with that of the instantaneous harmonic oscillator at that instant 
of time. 
For a single time-dependent oscillator, this procedure has been used in \cite{tapo1} to find out the
Nielsen complexity of the Lipkin-Meshkov-Glick model using the wave function method of \cite{Myers1},\cite{ABHKM}.
Recently, in \cite{CS} the solution of the EMP equation was also used to find out the complexity using 
the covariance matrix of a state and its comparison with the wavefunction method was also studied.

\subsection{Nielsen complexity with the state at initial time as the reference}

The quench protocol we consider  is the following. 
We consider the chain containing $N$ interacting 
oscillators with frequency $\omega_i$ and interaction strength $k_i$, and at time 
$t=0$ we change these parameters to different constant  values $\omega_f$ and $k_f$ respectively. 
Our goal here is to find out the circuit complexity of the state at an arbitrary time $t>0$
by taking the state at $t=0$ to be the reference state. This reference state, being the product of  
ground  states of $N$ oscillators of frequency $\sqrt{\lambda_{j(in)}}=\sqrt{\lambda_j(t=0)}$,
is also a Gaussian in the $y_j$ coordinates, just like the target state. In such a case, the expression 
of the Nielsen complexity is well known and is given by \cite{ABHKM}
\begin{equation}\label{complexity}
\mathcal{C}(t)=\frac{1}{2}\sqrt{\sum_{j=1}^{N}\Big(\mathcal{A}_j^2(t)+\mathcal{B}_j^2(t)\Big)}~,~
\mathcal{A}_j(t)=\ln\Bigg(\frac{|\Omega_j|}{\sqrt{\lambda_j(0)}}\Bigg)~,~
\mathcal{B}_j(t)=\arctan\bigg(\frac{\text{Im}(\Omega_j)}{\text{Re}(\Omega_j)}\bigg)~.
\end{equation}
With the form of $\Omega_j(t)$ given in Eq.  (\ref{groundstate}), the expressions for the time dependent functions 
$\mathcal{A}_j(t)$ and $\mathcal{B}_j(t)$ can be written directly in terms of
the auxiliary function $b_j(t)$ and its time derivative  as \cite{tapo1}
\begin{equation}\label{functions_bbdot}
\mathcal{A}_j(t)=\ln\Bigg[\frac{\sqrt{b_j^2(t)\dot{b}_j^2(t)+\lambda_j(0)}}
{\sqrt{\lambda_j(0)}b_j^2(t)}\Bigg]~,~\mathcal{B}_j(t)=\arctan\bigg(
\frac{b_j(t)\dot{b}_j(t)}{\sqrt{\lambda_j(0)}}\bigg)~.
\end{equation}

Now to obtain these functions for the quench process considered here,
we need the explicit solution of the auxiliary equation which satisfies the initial conditions of Eq.  (\ref{initial}).
To obtain the general solution of Eq. (\ref{auxiliary_eq}) with $\lambda_j(t)$ given above we follow the following 
procedure. We consider two linearly independent solutions of the classical equation of motion of the harmonic oscillator
with frequency $\lambda_j$
\begin{equation}\label{classical}
\frac{d^{2}g_j(t)}{dt^{2}}+\lambda_j g_j(t)=0~,
\end{equation}
which we denote respectively as $g_{j1}(t)$ and $g_{j2}(t)$. Then a general solution of the corresponding auxiliary  
equation can be written as 
\begin{equation}
b_j(t)=\sqrt{A_{j}g_{j1}^2(t)+2B_{j}g_{j1}(t)g_{j2}(t)+C_{j}g_{j2}^2(t)}~,
\end{equation}
with the three consonants $A_{j},B_{j}, C_{j}$ are being related by the relation
\begin{equation}\label{independent_cons}
A_{j}C_{j}-B_{j}^2=\pm\frac{\lambda_j(0)}{W^2}~,
\end{equation}
where $W$ the Wronskian of  the two linearly independent classical solution. Any two independent 
constants, say, $A_{j}$ and $C_{j}$ can be obtained by using the conditions listed in 
Eq. (\ref{initial}) at $t=0$. 

For our case, since after the quench at $t=0$ the frequency $\sqrt{\lambda_j}$ takes a
constant value, the solution for the auxiliary equation can be written with $y=\sqrt{\lambda_j}t$ as
\begin{equation}
b_{j}(t)=\sqrt{A_j\cos^2 y + 2B_j\sin y\cos y
+C_j\sin^2 y} = \sqrt{\alpha_j\cos(2 y)+\beta_j\sin(2 y)+\gamma_j}~.
\end{equation}
The relation between the two sets of constants can be worked out easily. Now, using the initial conditions
and the relation of Eq. (\ref{independent_cons}) between the three constants,
we can obtain these constants to be 
\begin{equation}
\alpha_j=\frac{\lambda_j-\lambda_j(0)}{2\lambda_j}~,~~\beta_j=0~,
~\gamma_j=\frac{\lambda_j+\lambda_j(0)}{2\lambda_j}~.
\end{equation}
With this solution for the auxiliary function the exact expressions for the functions $\mathcal{A}_j(t)$
and $\mathcal{B}_j(t)$ are given by
\begin{equation}\label{functions}
\mathcal{A}_j(t)=\ln\Bigg[\frac{\sqrt{\alpha_j^2\lambda_j\sin^2(2y)+\lambda_j(0)}}
{\Big(\gamma_j+\alpha_j\cos(2y)\Big)\sqrt{\lambda_j(0)}}\Bigg]~,
~\mathcal{B}_j(t)=\arctan\bigg[\frac{\alpha_j\sqrt{\lambda_j}\sin(2y)}
{\sqrt{\lambda_j(0)}}\bigg]~.
\end{equation}
Putting these back in Eq. (\ref{complexity}), we obtain the time evolution of the circuit complexity.
Before studying this evolution we consider the following special cases.

\subsection{Early time growth and the zero mode contribution}

We first study the behaviour of the complexity at $t\rightarrow0$, i.e., just after the quench. 
The behaviour of other information theoretic quantities such as the entanglement entropy have been 
studied previously at this limit. 
From the  expressions in Eq. (\ref{functions}), we see that at $t=0$,
both $\mathcal{A}_j(t=0)=0$, $\mathcal{B}_j(t=0)=0$, hence at $t=0$ we have  
$\mathcal{C}(t=0)=0$. Higher order terms in the expansion can be obtained  using Eqs. (\ref{complexity})
and (\ref{functions_bbdot}) and are given as $\mathcal{C}^2(t)=a_2 t^2+a_4t^4+\cdots $, with
\begin{equation}\label{expansion_in_t}
a_2=\frac{1}{4}\sum_{j=1}^{N}\frac{\big(\lambda_j-\lambda_j(0)\big)^2}{\lambda_j(0)}~,
~a_4=\frac{1}{4}\sum_{j=1}^{N}\frac{\big(\lambda_j-\lambda_j(0)\big)^2\Big(5\lambda_j^2
-6\lambda_j\lambda_j(0)+5\lambda_j(0)^2\Big)}{12\lambda_j(0)^2}~,~\cdots ~.
\end{equation}
We note that, similar to the result obtained using the covariance matrices \cite{dGT1}, the complexity
squared only contains the even powers of time. Thus very close to the initial time, just after
the quench, the complexity grows linearly in time (here $a_2>0$).

As an alternative limit, we can also calculate the first order contribution to the complexity 
when the target frequency is only infinitesimally different from the reference frequency, i.e.,
$\omega_{f}=\omega_i+\delta$, with $\delta\ll1$. In this case the $\mathcal{O}(\delta)$
contributions to the  functions $\mathcal{A}_j$ and $\mathcal{B}_j$ with $k_f=k_i$ are given by
\begin{equation}
\mathcal{A}_j(\omega_{f}=\omega_i+\delta)\approx 
\frac{2\omega_i\sin^2(\sqrt{\lambda_j(0)}t)}{\sqrt{\lambda_j(0)}}\delta+\mathcal{O}(\delta^2)~,~\\
\mathcal{B}_j(\omega_{f}=\omega_i+\delta)\approx  
-\frac{\omega_i\sin(2t\sqrt{\lambda_j(0)})}{\sqrt{\lambda_j(0)}}\delta+\mathcal{O}(\delta^2)~.
\end{equation}
These two terms contribute at $\mathcal{O}(\delta)$ to the complexity. 
It can now be seen that when we take the reference frequency to be zero, the complexity does not
have any $\mathcal{O}(\delta)$ contribution. This is in contradiction with the results obtained 
from the covariance method, where there is an $\mathcal{O}(\delta)$ contribution to the complexity in this case.
This result indicates that for small change of the reference frequency
from the initial zero value, the complexity calculated from the wavefunction cannot detect the 
lowest order change in the complexity, whereas complexity computed using the covariance matrix 
shows a linear behaviour, one that is of first order in the target state frequency.
	
Next, we investigate the contribution of the $N$-th mode towards the total complexity.
As we shall see below, the $N$-th mode contribution determines the behaviour of the complexity at late times. For the 
complexity calculated using the covariance matrix method, the contribution of the $N$-th mode has been
investigated in details in \cite{dGT1}. When $\lambda_j\to 0$, this is the zero mode contribution.
Here, our goal is to analyse this contribution with our formula.

The contribution of $N$-th mode (denoted by $\mathcal{C}_{0}$) and the rest can be written as
$\mathcal{C}^2(t)=\mathcal{C}_{r}^2(t)+\mathcal{C}_{0}^2(t)$, with 
$\mathcal{C}_{r}^2(t)=\frac{1}{4}\sum_{j=1}^{N-1}\Big(\mathcal{A}_j^2(t)+\mathcal{B}_j^2(t)\Big)$, 
and $\mathcal{C}_{0}^2=\frac{1}{4}\Big(\mathcal{A}_{N}^2+\mathcal{B}_N^2 \Big)$, i.e.,
\begin{equation}\label{lowerbound}
\mathcal{C}_{0}^2=\frac{1}{4}\bigg(\ln\Bigg[\frac{\sqrt{\alpha_N^2\lambda_N\sin^2(2y_N)+\lambda_N(0)}}
{\Big(\gamma_N+\alpha_N\cos(2y_N)\Big)\sqrt{\lambda_N(0)}}\Bigg]\bigg)^2 +\frac{1}{4}
\bigg(\arctan\bigg[\frac{\alpha_N\sqrt{\lambda_N}\sin(2y_N)}
{\sqrt{\lambda_N(0)}}\bigg]\bigg) ^2~,
\end{equation}
where we have denoted $y_N=\sqrt{\lambda_N}t$.
With this expression for the $N$-th mode contribution, we can provide an upper 
and lower bound to the circuit complexity as $\mathcal{C}_0(t)\leq\mathcal{C}(t)\leq\mathcal{C}_{u}$, with
\begin{equation}\label{upperbound}
\mathcal{C}_u^2(t)=\mathcal{C}_0^2(t)+\frac{1}{4}\sum_{j=1}^{N-1}
\Big[\mathcal{A}_{uj}^2+\mathcal{B}_{uj}^2\Big]~,~
\mathcal{A}_{uj}=\ln\Bigg[\frac{\sqrt{\alpha_j^2\lambda_j+\lambda_j(0)}}
{\gamma_j\sqrt{\lambda_j(0)}}\Bigg]~,~\mathcal{B}_{uj}
=\arctan\bigg[\frac{\alpha_j\sqrt{\lambda_j}}{\sqrt{\lambda_j(0)}}\bigg]~,
\end{equation}
These bounds will be plotted along the complexities in the plot below to check their validity. 

	
\begin{figure}[h!]
\begin{minipage}[b]{0.45\linewidth}
\centering
\includegraphics[width=0.95\textwidth]{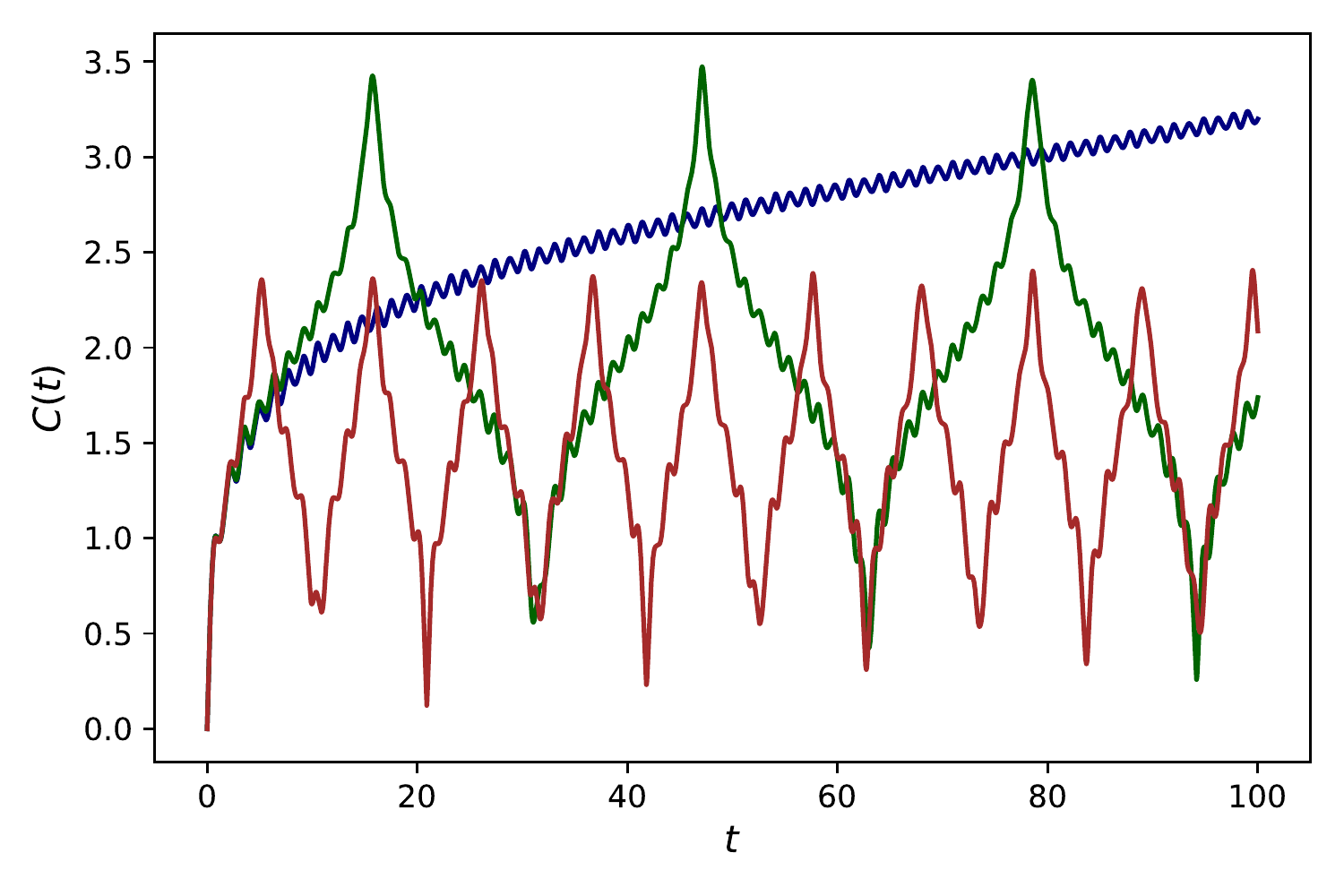}
\caption{Time evolution of complexity after a single quench, with the same reference state 
frequency and different target state frequencies and $N=4$. The red, green and blue 
curves are for $\omega_f=0.3, 0.1, 0.01$, with $k_i=2, k_f=2.5,\omega_i=3$. 
Dynamically generated time scales can be seen from this plot.}
\label{fig:C_with_N=4 }
\end{minipage}
\hspace{0.2cm}
\begin{minipage}[b]{0.45\linewidth}
\centering
\includegraphics[width=0.95\textwidth]{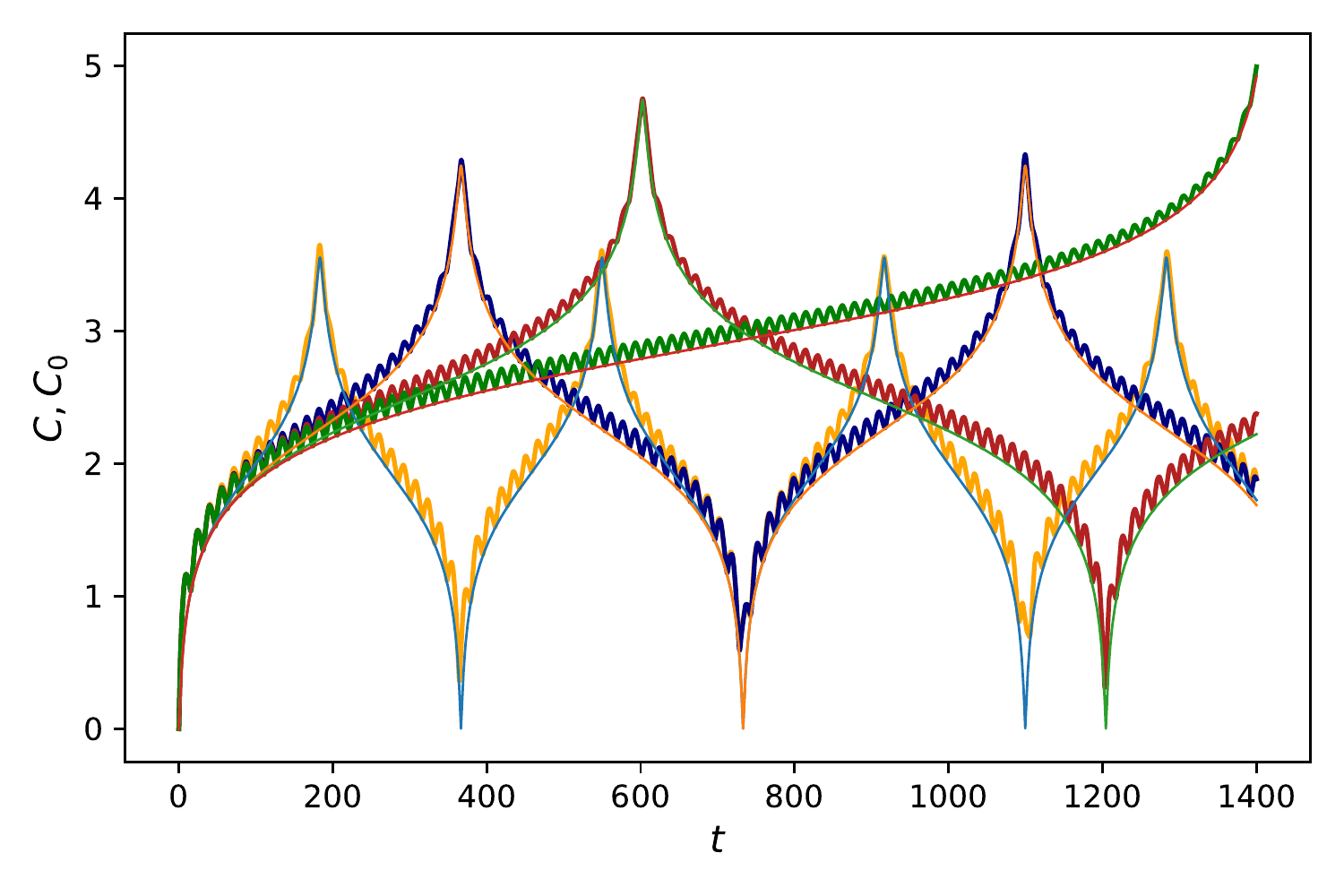}
\caption{Time evolution of complexity after a single quench with same reference state 
frequency and different target state frequencies and $N=100$.
Here, $\omega_i=0.3$, $k_f=k_i=10$. The yellow, blue, red and green curves denote
$\omega_{f}=0.009, 0.004$, 0.002$, 0.001$. The lower bound in Eq. (\ref{lowerbound}) for 
each target frequency is also plotted.}
\label{fig:C_with_sin_qu1}
\end{minipage}
\end{figure}
	

\subsection{Time evolution of complexity : revivals}
	
The time evolution of the complexity after a quench is shown in Fig. (\ref{fig:C_with_N=4 }) when 
the harmonic chain contains four interacting oscillators, whose interaction strength changes from 
$k_i=2$ to $k_f=2.5$ and the frequency $\omega_i=3$ is changed to $\omega_f=0.3,0.1$ and $0.01$.
To make a  comparison with  plots of von Neumann entropy, we have kept the parameter values the same as in 
reference \cite{GGS1}. Very similar to the plots of the von Neumann entropy in that work, circuit complexity show 
revivals during the evolution, and the time period of revival increases as the values of the post-quench
frequency is decreased. Furthermore, similar to the von Neumann entropy, complexity also shows
quasi-revivals of shorter time period, with these time periods generated dynamically.
Furthermore, neither the entanglement entropy, nor the complexity saturates with time. 

There are however, important differences between them as well. 
Foe example there are sharp peaks  present in the complexity, with the  magnitude of theses
peaks increase with decreasing the value of the final frequency, i.e., as we move away from the initial frequency.
These peaks are markedly absent from the plots of the von Neumann  entropy.
It  can be checked that in the expression of Eq. (\ref{complexity})
the term $\sum_{j=1}^{n}\mathcal{A}_j^2$ is responsible for these peaks.
On the space of unitary operators these points represent the maximum distance between the target state 
and the reference state and it is determined by the magnitude of the reference and the target state frequencies   
rather than their complex phase factors. Furthermore, though complexity has sharp minima, it never reaches zero,
i.e., the time evolved state never coincides with the state before quench at $t=0$.   
	
The time evolution of complexity with a large number of particle is shown in Fig. \ref{fig:C_with_sin_qu1},
with particle number $N=100$. Once again, the plots of the complexity show sharp peaks besides the 
presence of the dynamically generated time periods of oscillation as well as quasi-revivals. Revivals 
observed here for the complexity calculated using the wavefunction method is also shared by the complexity
obtained rom the covariance matrix \cite{dGT1}.

In the same figure, we have also plotted the corresponding lower bounds of the complexity given in Eq. (\ref{lowerbound}).
For all the target state frequencies, the lower bounds are very useful. The upper bound to the complexity 
given in Eq.  (\ref{upperbound}) can also be shown to provide excellent consistency, however we have 
not shown it here, in order not to clutter the diagram. 

\subsection{Time evolution after critical quench}\label{critical_quench}

Next we shall consider the time evolution of complexity after a critical quench.
For the critical evolution, the frequency after the quench vanishes i.e. $\omega_{f}=0$. In this limit,
the zero mode contribution can be simplified as
\begin{equation}\label{lowerbond_critical}
\mathcal{C}_{0c}^2=\frac{1}{16}\Big(\ln\big[1+t^2\omega_i^2\big]\Big)^2+
\frac{1}{4}\Big(\arctan\big[\omega_it\big]\Big)^2~.
\end{equation}
It can now be seen that, due to presence of the first term in the zero mode, at  
$t\rightarrow\infty$, the complexity diverges logarithmically (the second term takes
a constant value in this limit).  

\begin{figure}[h!]
\begin{minipage}[b]{0.45\linewidth}
\centering
\includegraphics[width=0.95\textwidth]{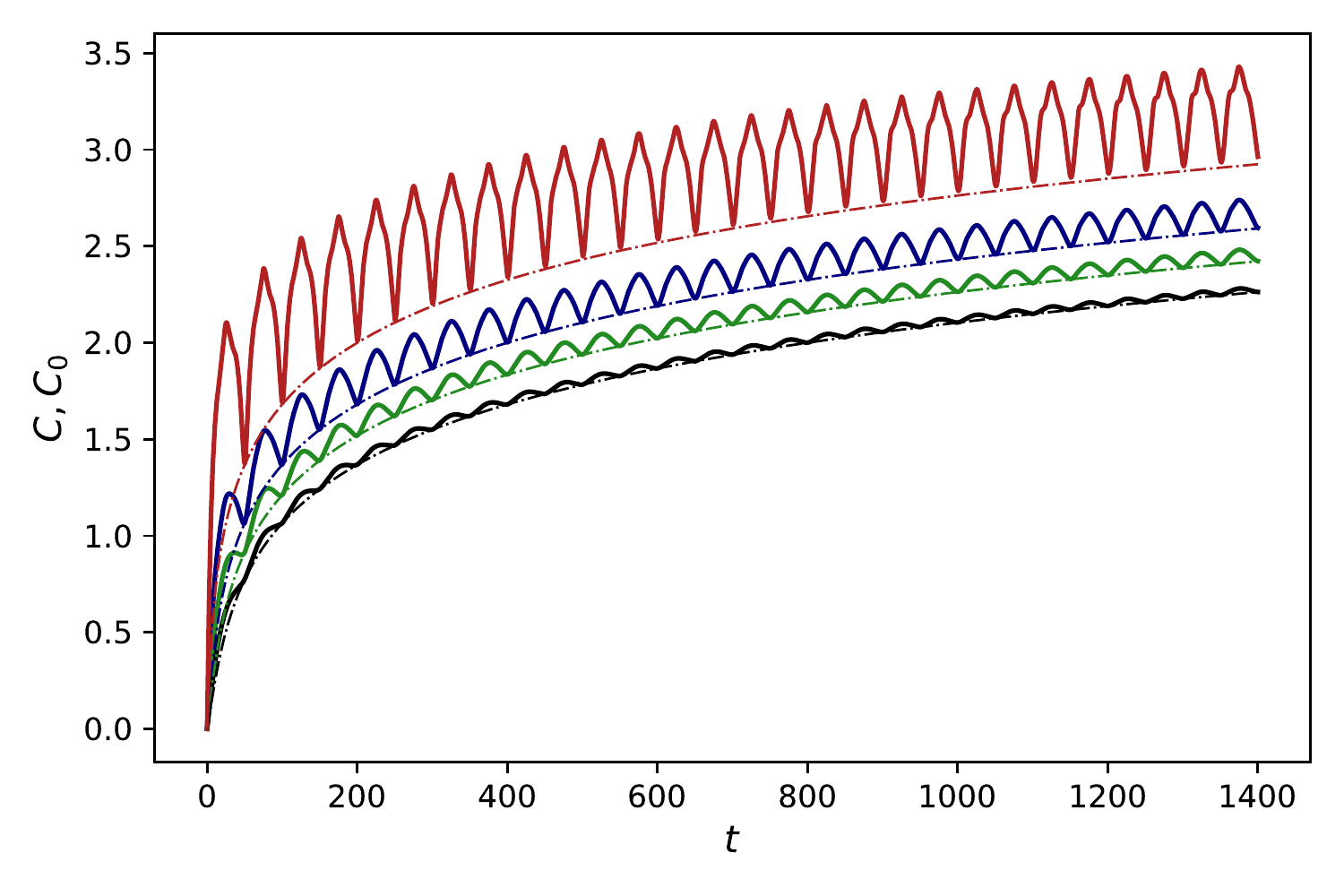}
\caption{Time evolution of complexity of after the critical quench with $N=100$. The black,
green, blue and red lines correspond to $\omega_i=0.05$, 0.07$, 0.1, 0.2$, with
$k_i= k_f=1$ and $\omega_f=0$.
The lower bound in this case in Eq. (\ref{lowerbond_critical}) is also shown  for each reference frequency.}
\label{fig:Critical_quench}
\end{minipage}
\hspace{0.2cm}
\begin{minipage}[b]{0.45\linewidth}
\centering
\includegraphics[width=0.95\textwidth]{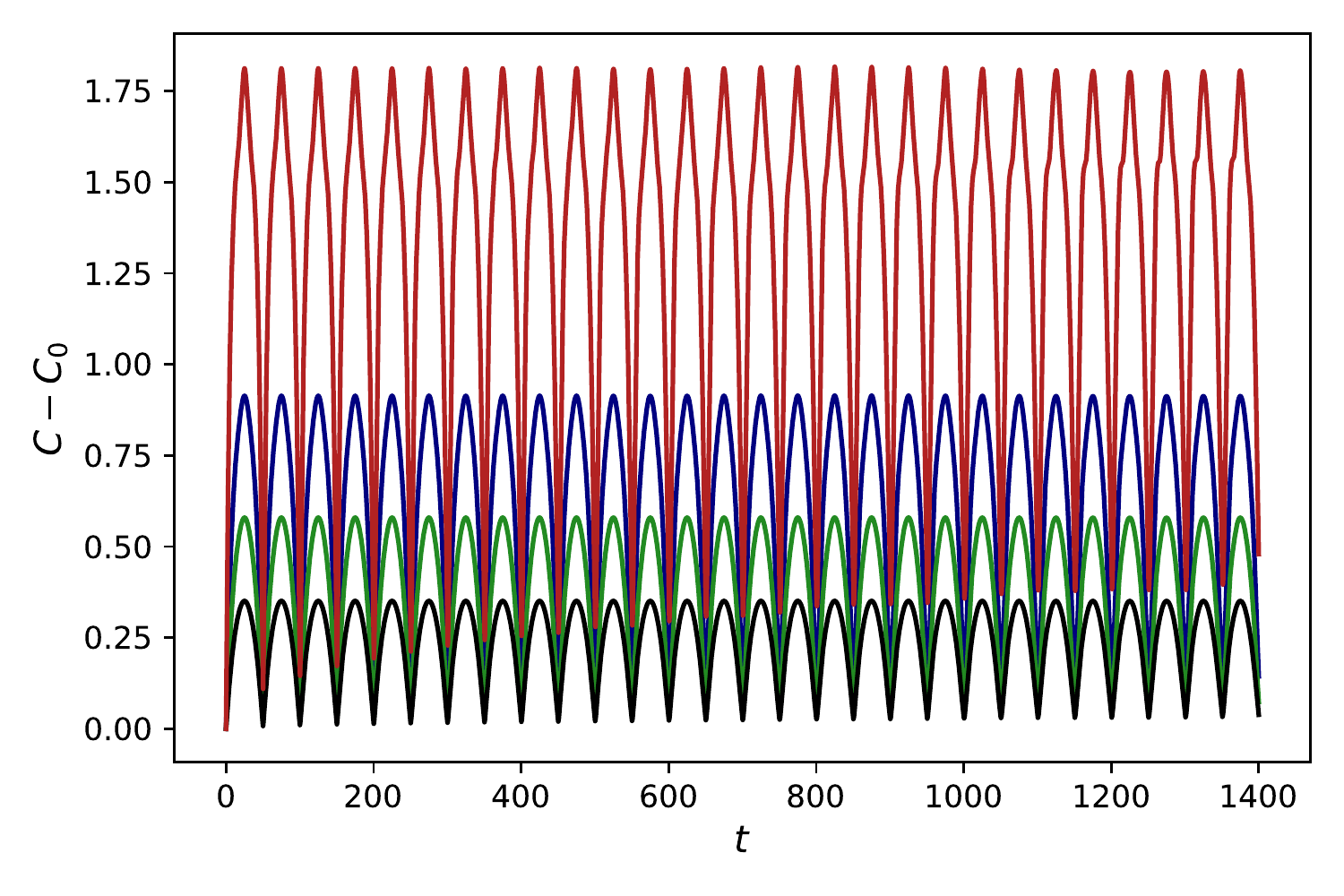}
\caption{Time evolution of complexity minus the zero mode contribution for critical quench with the
parameter value same as Fig. \ref{fig:Critical_quench}, with the same colour coding. 
The zero mode is responsible for growth of complexity with time. }
\label{fig:C_critical_minius_zero}
\end{minipage}
\end{figure}

Time evolution of circuit complexity after the critical quench for different values of the reference state
frequencies are plotted in Fig. \ref{fig:Critical_quench} with $N=100$. 
The lower bounds in Eq. (\ref{lowerbond_critical}) for different reference state are also plotted in the same figure.
With increasing reference state frequency, as well as for large times, this bound becomes slightly less reliable.
As can be seen from these plots, in contrast to the non-critical cases,
for the critical quench, complexity grows with time. This is due to the divergence of the 
zero mode contribution at large times. To see this more explicitly we have plotted in Fig. \ref{fig:C_critical_minius_zero},
corresponding to the complexities plotted in Fig. \ref{fig:Critical_quench} the total complexity minus
the zero mode contribution.
When the zero mode is subtracted, the complexity does not increase, and rather oscillates with time.

The growth of the circuit complexity with time for the critical quench is also observed in the 
complexity calculated using the covariance matrices in \cite{dGT1}. However, in this case, when the 
zero mode contribution is subtracted from the complexity the resulting expression shows decoherence with
time for relatively higher values of the reference state frequency. This is in sharp contrast with our case,
where the complexity calculated using the wavefunction does not show such decoherence with time. 

\section{Complexity evolution in a harmonic chain under multiple quenches}

Now we will investigate the time evolution of circuit complexity in the harmonic chain of the 
previous section, when it is subjected successively to more than one quench.
For such multiple quench protocol, the time evolution of entanglement entropy and the 
out of time order correlator (OTOC) have been recently studied in \cite{GGS2}, where 
marked changes of these quantities from their single quench counterpart are observed.

Our quench protocol is the following.
At $t=0$, the frequency $\omega_i$ is changed to $\omega_f$ by keeping the other parameter $k$ unchanged. 
Now, after a time period $T$, the new frequency $\omega_f$ is changed back to the old one $\omega_i$, thereby 
defining a second quench. This sequence is now repeated to define a multiple quench protocol.
For now we shall assume that all the quenches are non-critical in nature. The case when any quench of the sequence is
critical will be dealt with later.

The solution for the auxiliary equation for the $j$th particle after the $i$th quench of time period $T$
can be written in terms of $y=\sqrt{\lambda_j}t$ as 
\begin{equation}\label{general_solution}
b_{j,i}\big((i-1)T<t<iT\big)=\sqrt{A_{j,i}\cos^2 y+2B_{j,i}\sin y
\cos y +C_{j,i}\sin^2 y}
=\sqrt{\alpha_{j,i}\cos(2y)
+\beta_{j,i}\sin(2y)+\gamma_{j,i}}~.
\end{equation}
Here each of these solutions are valid for their respective range of time period of quench,
and hence we need to use the appropriate expression for $\lambda_j$, either
$\lambda_{(f)j}(t)=\omega^2_f(t)+2k(t)\big[1-\cos(\frac{2\pi j}{N})\big]$, or 
$\lambda_{(i)j}(t)=\omega^2_i(t)+2k(t)\big[1-\cos(\frac{2\pi j}{N})\big]$, 
depending on the time period under consideration.
To determine the constants appearing in the above solutions, we demand that the solutions $b_{j,i}(t)$
and their time derivative $ \dot{b}_{j,i}(t)$ are continuous for each value of the index 
$i$ i.e., after and before a quench their values should match.
This two conditions determine two of the above constants and the 
third can be determined by using the relation (\ref{independent_cons}) between them.\footnote{The relations 
corresponding to that of Eq. \ref{independent_cons} for the second set of 
constants are given by $\gamma_{j,i}^2-\beta_{j,i}^2-\alpha_{j,i}^2= \lambda_j(0)/\lambda_{(i)j}$
or $\gamma_{j,i}^2-\beta_{j,i}^2-\alpha_{j,i}^2= \lambda_j(0)/\lambda_{(f)j}$ depending on the value of time $t$.}
Furthermore, at $t=0$, the conditions at Eq. (\ref{initial}) are valid as well, which (along with Eq. (\ref{independent_cons}))
should determine the constants $A_{j,1},B_{j,1},C_{j,1}$. The expressions for the rest of the constants can 
be obtained in terms of them.

Let us suppose that we have performed $n$ quenches following the protocol discussed above. 
First, we shall study the time evolution of complexity between a reference state at $t=0$ and a 
target state at a time $t$ after the $n$th quench. The formula for the complexity $\mathcal{C}_i$
between the time interval $(i-1)T<t<iT$ can be written as
\begin{equation}\label{complexity_multipile}
\mathcal{C}_i^2\big((i-1)T<t<iT\big)
=\frac{1}{4}\sum_{j=1}^{N}\Big(\mathcal{A}_{j,i}^2(b_{j,i},
\dot{b}_{j,i})+\mathcal{B}_{j,i}^2(b_{j,i},\dot{b}_{j,i})\Big)~,
\end{equation} 
where we have defined
\begin{equation}
\mathcal{A}_{j,i}(t)=\ln\Bigg[\frac{\sqrt{b_{j,i}^2(t)\dot{b}_{j,i}^2(t)+\lambda_{(i/f)j}}}
{\sqrt{\lambda_{(i)j}}b_{j,i}^2(t)}\Bigg]
\quad \text{and}\quad \mathcal{B}_{j,i}(t)=\arctan\bigg[
\frac{b_{j,i}(t)\dot{b}_{j,i}(t)}{\sqrt{\lambda_{(i)j}}}\bigg]~.
\end{equation}
Here the notation $\lambda_{(i/f)j}$ represents $\lambda_(i)j$ or $\lambda_{(f)j}$
depending on the quench number.
Thus the total complexity contains $n$ different parts corresponding to the individual
complexities in  each section of the quench. Since the quantities $\mathcal{C}_i$
depend on auxiliary functions $b_{j,i}$ and their derivatives, and they are continuous
between successive quenches, the complexity at an arbitrary time $t$ is a continuous function.

For the solutions of the auxiliary functions written above, 
these functions are given by
\begin{equation}\label{functions_multiple}
\begin{split}
\mathcal{A}_{j,i}(t)=&\ln\Bigg[\frac{\sqrt{\lambda_{(i)j}+\big(\beta_{j,i}\cos(2\sqrt{\lambda_{(i/f)j}}t)
-\alpha_{j,i}\sin(2\sqrt{\lambda_{(i/f)j}}t)\big)^2\lambda_{(i/f)j}}}
{\Big(\alpha_{j,i}\cos(2\sqrt{\lambda_{(i/f)j}}t)+\beta_{j,i}\sin(2\sqrt{\lambda_{(i/f)j}}t)+\gamma_{j,i}\Big)
\sqrt{\lambda_{(i)j}}}\Bigg]~,\\
\mathcal{B}_{j,i}(t)=&\arctan\bigg[\frac{\Big(\alpha_{j,i}\sin(2\sqrt{\lambda_{(i/f)j}}t)
-\beta_{j,i}\cos(2\sqrt{\lambda_{(i/f)j}}t)\Big)\sqrt{\lambda_{(i/f)j}}}{\sqrt{\lambda_{(i)j}}}\bigg]~.
\end{split}
\end{equation}
It can be easily seen that  just after the first quench, for a time $t<<T$ the complexity grows proportional
to $t$, since in this case the evolution of complexity is the same as in a single quench. 
Similarly, after each quench we can expand the complexity in powers of $t$, and we
obtain the following series (with the time being shifted for the higher quench number)
$\mathcal{C}_i^2(t)=a_{i0}+a_{i1}t+a_{i2}t^2+\cdots$, where now the complexity has a finite value at 
zeroth order in time, given by
\begin{equation}
\label{muliple_expansion}
a_{i0}=\frac{1}{4}\sum_{j=1}^{N}\Bigg\{\bigg(\log\Bigg[\frac{\big(\alpha_{j,i}+\gamma_{j,i}
\big)\sqrt{\lambda_{(i)j}}}{\big(1+\beta_{j,i}^2\big)
\sqrt{\lambda_{(i/f)j}}}\Bigg]\bigg)^2+\Bigg(\arctan \bigg[
\frac{\beta_{j,i}\sqrt{\lambda_{(i/f)j}}}{\sqrt{\lambda_{(i)j}}}\bigg]\Bigg)^2\Bigg\}~.
\end{equation}
When we consider the first quench this quantity will be zero. In contrast to the 
single quench  the linear growth is now determined by the coefficient of $t$ ($a_{i1}$) 
and the zeroth order contribution ($a_{i0}$).
	
The zero mode contribution for the multiple quench now consists of $n$ different contributions  
accounting for $n$ quenches and the contribution of the $i$th 
quench can be written as follows
\begin{equation}
\mathcal{C}^2_{i0}=\frac{1}{4}\Big(\mathcal{A}_{N,i}^2(b_{N,i},
\dot{b}_{N,i})+\mathcal{B}_{N,i}^2(b_{N,i},\dot{b}_{N,i})\Big)~.
\end{equation}
The expressions for the functions $\mathcal{A}_{N,i}^2(b_{N,i},\dot{b}_{N,i})$ and
$\mathcal{B}_{N,i}^2(b_{N,i},\dot{b}_{N,i})$ can be read from Eqs. (\ref{functions_multiple}) above.
During the complexity evolution after each quench these provide a lower bound to the complexity
in the time period under consideration.
	
\subsection{Complexity evolution with multiple quenches}

Now we shall study the time evolution of the complexity between a reference state at $t=0$ 
and a target state  at an arbitrary time $t$ numerically   by using the formula in Eq. 
(\ref{complexity_multipile}) which is valid when the target state is between $(i-1)T
<t<iT$. Here we impose a smooth matching between the auxiliary functions and its time derivative at times $iT$,
with $i=1,2..n$ i.e. after and before any particular quench.
The values of the first set of constants $\alpha_{j,1},\beta_{j,1},\gamma_{j,1}$
can be obtained by imposing the conditions of Eq.  (\ref{initial}) at $t=0$,
and they are given by: $\alpha_{j,1}=\frac{\lambda_{(f)j}-\lambda_{(i)j}}{\lambda_{(f)j}}, 
\beta_{j,1}=0,\gamma_{j,1}=\frac{\lambda_{(f)j}+\lambda_{(i)j}}{\lambda_{(f)j}}$.
By using these values, the continuity of the auxiliary functions and its derivative across
each quench and the relation between the three constants mentioned above the other 
independent constants can be obtained at successive quenches. 
	
\begin{figure}[h!]
\begin{minipage}[b]{0.45\linewidth}
\centering
\includegraphics[width=0.95\textwidth]{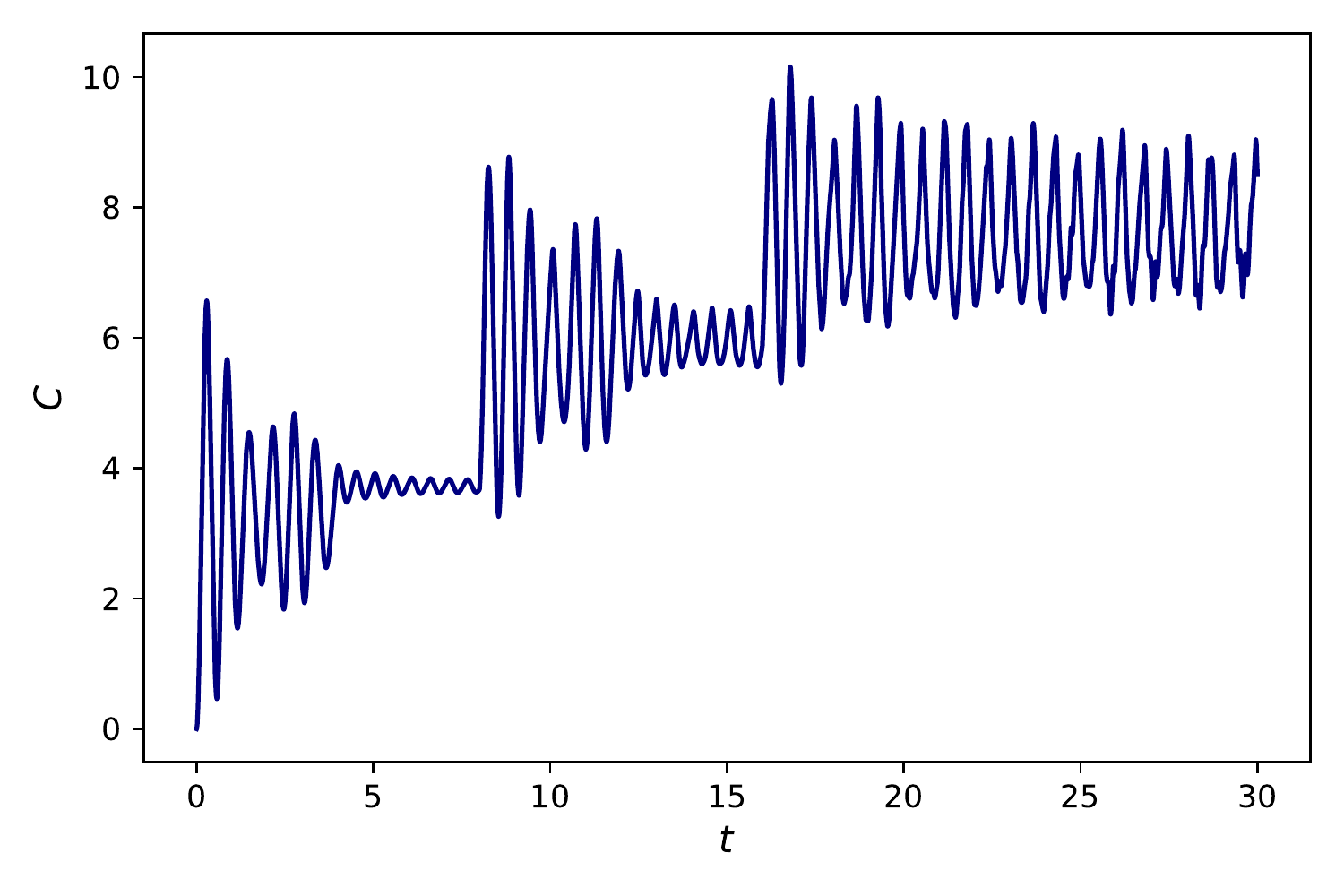}
\caption{Time evolution of complexity with five successive quenches
with $N=100$. Here, $k_f=k_i=4,T=4, \omega_f=5$ and $\omega_i=3$.}
\label{fig:Complexity_3quenchs}
\end{minipage}
\hspace{0.2cm}
\begin{minipage}[b]{0.45\linewidth}
\centering
\includegraphics[width=0.95\textwidth]{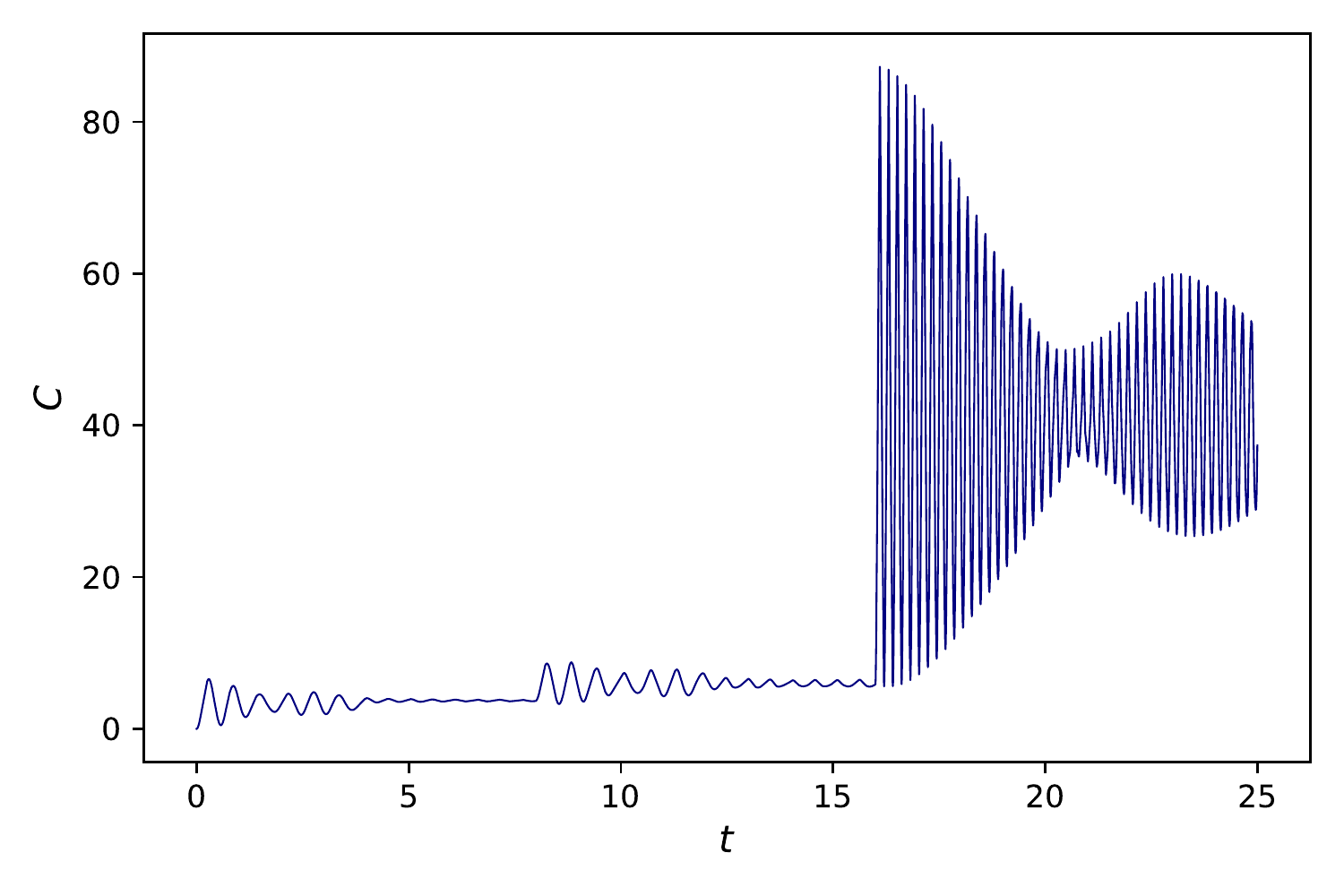}
\caption{Time evolution of complexity with five 
successive quenches when the final quench frequency is
much higher than the preceding ones. Here, we 
set $\omega_f=5$ after first and third 
quench, and $\omega_{i}=3$ after second and fourth quench.  For fifth
quench frequency is $\omega_{f}=15$. }
\label{fig:C_multiple_large_frequncy}
\end{minipage}
\end{figure}

In Fig. \ref{fig:Complexity_3quenchs} we have plotted the time evolution of the circuit 
complexity after five successive quenches between frequency $\omega_i=3 $ and $\omega_{f}=5$,
with $k_f=k_i=4$, $T=4$, for a harmonic chain containing 100 particles.
The  complexity shows  oscillating behaviour, with the magnitude 
of the oscillations being  dependent  on the frequency of the chain.  From this plots we can 
make the following observations. \footnote{These conclusions are valid for frequency $\omega_{f}$ sufficiently
higher than $\omega_i$. We are not considering critical quenches in this section.}
	
Firstly, comparing the complexities of the states having the same frequencies $\omega_i$, we see that
among these equal frequency states the state at higher times has higher complexity.
Hence even after two successive quenches when we come back to the same frequency $\omega_i$  
 the unitary operator corresponding to this 
state does not come back to its previous position on the space of unitary operators,
rather it further moves away from it.  The reason for this behaviour is that, after two quenches,
even though  the frequency returns to its pre-quench value the solution for the auxiliary function,
and hence the wavefunction is necessarily different from their expressions at $t=0$ due to 
the inherent time dependence of the problem. Thus there is always some  `residual' complexity that is
present between two states at $t=0$ and $t\geq4$ due to the time dependence of the problem.
However, with two successive quenches between frequency $\omega_f$ the complexity does comes to values 
same as that of previous quench with the same frequency.

Secondly, comparing the complexities  of the time period of the $i$th quench with frequency $\omega_i$
and that of the $(i+1)$th quench, having frequency $\omega_f$ we notice that, for sufficiently high
value of $\omega_f$ compared with the initial frequency, there are times just after 
the $(i+1)$th quench when the complexity evolves to  values smaller even than those of the $i$th quench.
These points on the space of unitary operators are closer to the initial state than those of the
state with  $i$th reference frequency. Gradually these oscillations dry out slightly so that complexity
oscillates around a value higher than that of the previous quench. However even at large times these oscillations
do not dry out completely, so that complexity does not attain a steady state value.

In this context an interesting question is the following: since the complexity just after the $(i+1)$th
quench is smaller than the $i$th quench, and as we have seen
above the magnitude of the oscillations of the complexity depends on the frequency of
the quench is it possible to decrease the complexity after  the $(i+1)$th quench
to a value which is much lower than that of the $i$h quench. In particular is it
possible to  make the complexity very close to zero, i.e. to made the
time evolved state after multiple quenches approach the reference state at $t=0$
by increasing the magnitude of the quench frequency? 
We have shown the time evolution of complexity in such an example in Fig. \ref{fig:C_multiple_large_frequncy}, 
where the final quench frequency is much larger
than the other frequencies. It can be seen from this plot, the complexity
after the fifth quench does not 
get lower than a value much smaller than the fourth quench.
Thus there is limit how far the time evolved state after a particular quench can reach
close to the reference state. This minimum value can not be decreased even by increasing the quench frequency sufficiently. 

If we now increase the number of quenches the 
complexity after the final quench oscillates around increasingly higher values. Thus, after a  sufficiently large
number of quenches is applied to the system a target state can be prepared with arbitrary large complexity. This 
is in sharp contrast with the time evolution of the complexity after a single non-critical quench, where the maximum value 
of the complexity is given by sharp peaks which are fixed ones the targets state frequency is fixed.
Hence after the non-critical single quench the unitary operator corresponding to the target state can not go beyond 
a particular point in the space of unitary operators, whereas after successive quenches the unitary operator
can be at a large distance from the identity. 	
	
In this context it is important to quantify the differences between the entanglement entropy (EE)
of the bipartite system and the circuit complexity. After multiple quenches the complexity does 
not saturates to any constant  value, rather it continue to oscillates around a mean value even at late times.
This is in contrast with the behaviour of other information theoretic quantities such as the entanglement entropy.
After multiple quenches, compared to the  single quench, the entanglement entropy attains a larger steady state 
value, and higher number of quenches leads to  a smaller fluctuation \cite{GGS2}. For the complexity the higher
number of quenches can not reduce its fluctuation. 
On the other hand the non trivial nature of the system after multiple quenches can be also infrared from the fact 
that in single quench scenario, the CC for various systems is known 
to saturate after initial linear growth, which in some cases even faster compared to the saturation of EE. 
This is ones again in contrast with the case considered here.
	

\subsection{Evolution when the final quench is a critical quench}

Now we shall discuss a special case of the multiple quench protocol, namely, when the final $n$th quench is 
a critical quench discussed in the context of a single quench in section \ref{critical_quench}. As we have seen 
the complexity in this case grows with time due to the logarithmic divergence of the zero mode contribution
towards the total complexity.

\begin{figure}[h!]
\begin{minipage}[b]{0.45\linewidth}
\centering
\includegraphics[width=0.95\textwidth]{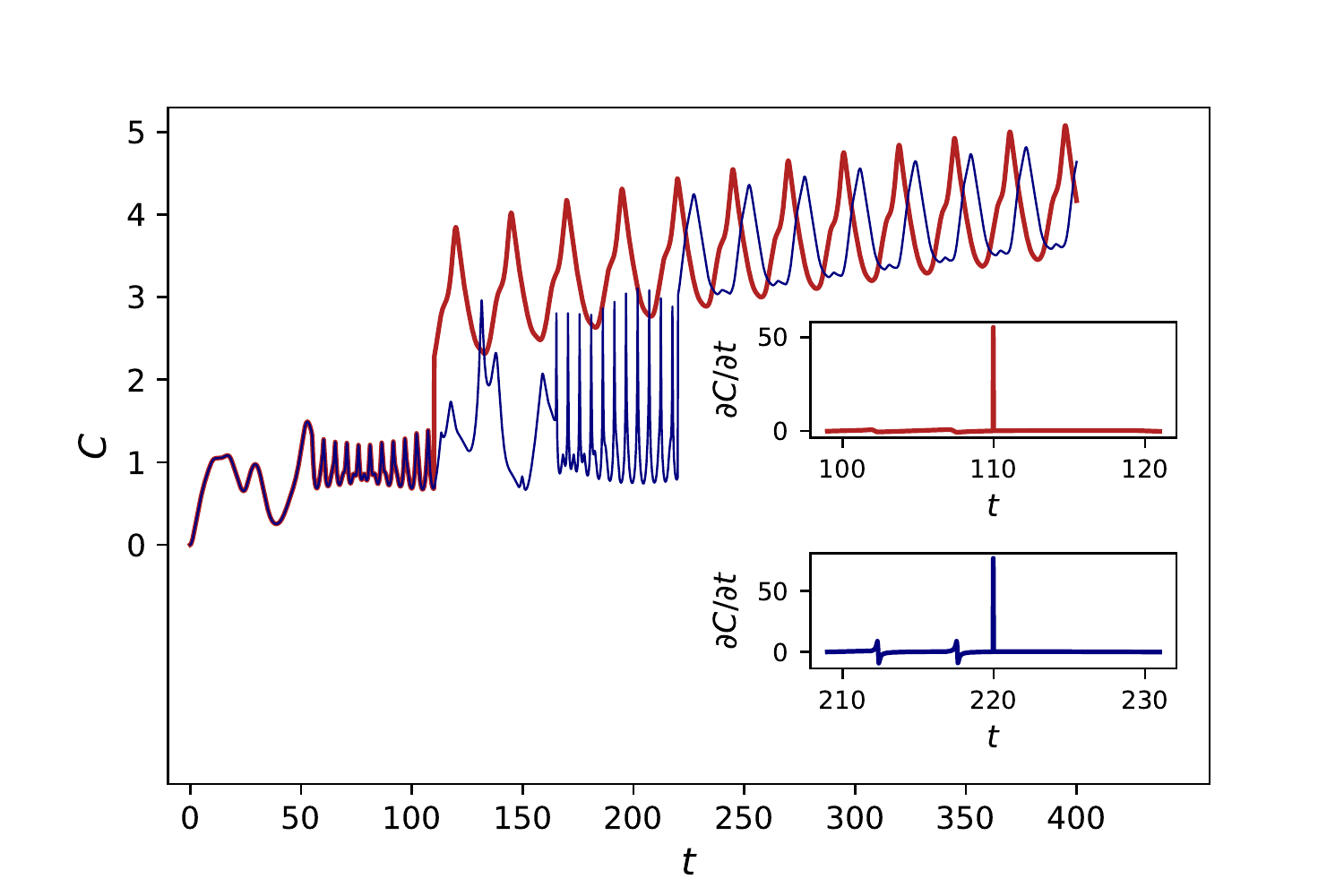}
\caption{Time evolution of complexity with three (Red) and five (Blue)
successive quenches when the final quench is critical quench. Here, $\omega_i=0.3, \omega_{f}=0.085$ (when non-critical),
$K_i=k_f=4$ and $T=55$. There is a discontinuity after the critical quench, which is evident from the 
plots of the derivatives of the complexity shown separately for two cases in the inset.}
\label{fig:C_critical_successive}
\end{minipage}
\hspace{0.2cm}
\begin{minipage}[b]{0.45\linewidth}
\centering
\includegraphics[width=0.95\textwidth]{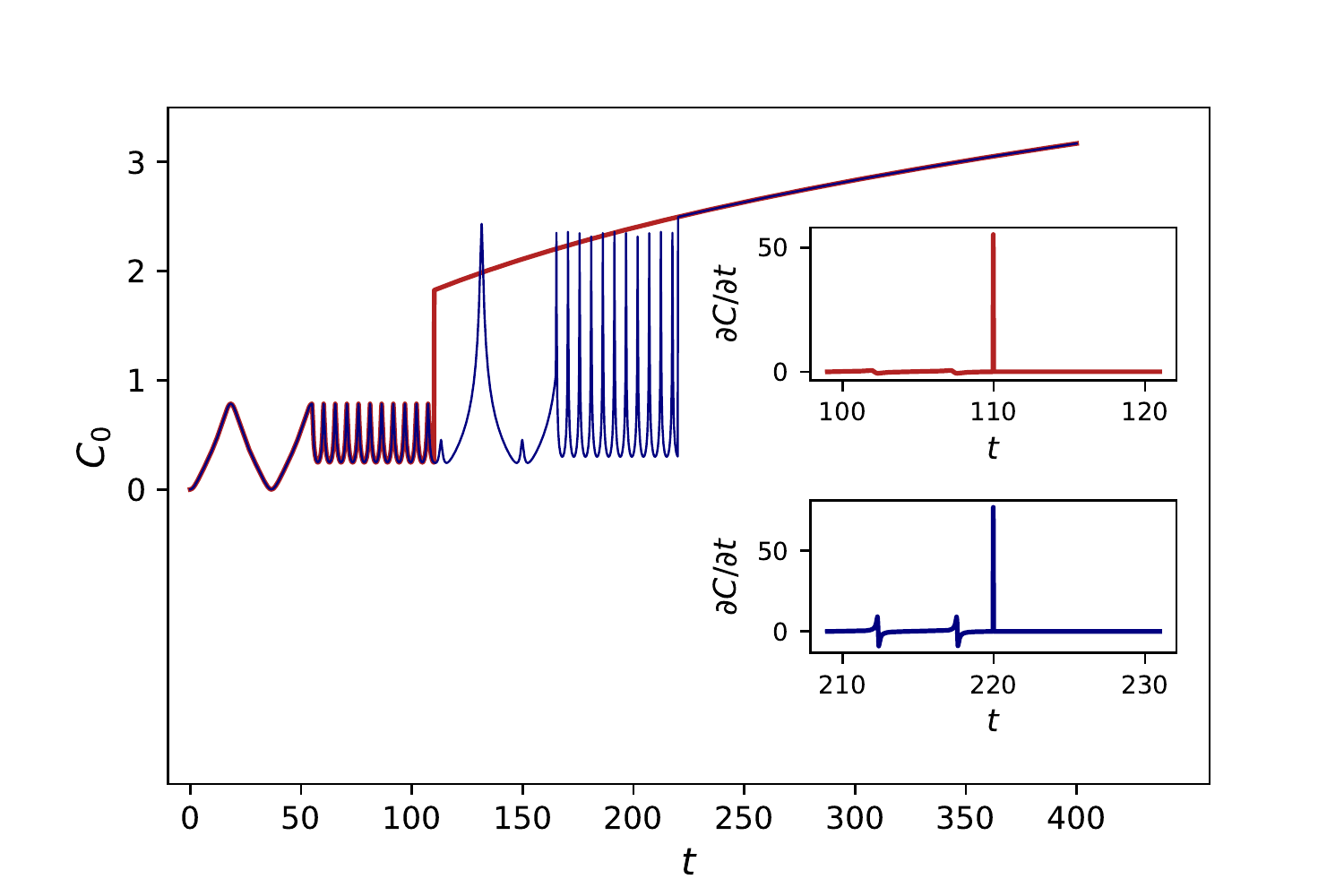}
\caption{$N$-th mode contributions corresponding to 
Fig. \ref{fig:C_critical_successive}.
There is a discontinuity after the critical quench.  The 
derivatives of the zero mode contribution are shown separately for two cases in the inset, from which the discontinuity is 
clearly seen.}
\label{fig:C_critical_successive_zero}
\end{minipage}
\end{figure}

In Fig. \ref{fig:C_critical_successive}, we have plotted the time evolution of the 
circuit complexity with three (red curve) and five quenches (blue curve) when the 
when the final quench is a critical quench. In both the plots after the characteristic oscillatory
behaviour of the complexity under the non-critical quench the the complexity
shows linear growth which is characteristic of the critical quench. However 
due to the fact that the final quench is a critical quench now 
there is a discontinuity in the complexity at the final quench which is markedly
absent from the non-critical multiple quench plots. This discontinuity  in the 
complexity is due to the discontinuity in the $N$-th mode contributions for
non-critical and critical quenches. As we have seen before at the limit of  critical
quench the zero mode contribution attains a particular value given by Eq. (\ref{lowerbond_critical}) which is different from the non-critical contributions
in Eq. (\ref{lowerbound}). 

To see this clearly we have plotted the $N$-th
mode contribution separately in Fig. \ref{fig:C_critical_successive_zero}, where
the discontinuity at the critical quench is visible in both the case of three and five quenches.

In both the Figures \ref{fig:C_critical_successive} and \ref{fig:C_critical_successive_zero}
the derivatives of the complexity and the zero mode contribution are also shown
separately which make the discontinuity of these quantities at the critical quench clearly visible.

\subsection{Complexity between states of subsequent quenches}
So far we have calculated the circuit complexity between states at $t=0$ and an arbitrary 
time evolved state after a single or multiple quenches. For more than one quenches we
can calculate  another set of quantities  which is not possible in the case
of a single quench, namely the complexity between states of successive quenches,
namely the complexity between a state before and after a particular quench, so that
the reference state is not the state at initial time.
These quantities will provide us with important information regarding the change of complexity after each quench
which are not captured when the reference state is a fixed state before the first quench.

As an example of calculation of such quantity
let us compute the circuit complexity between a reference state at a time $t=t_0$ 
of the $i$th quench and a target state at an arbitrary time $t$ between $t_0<t<(i+1)T$.
This quantity will help us to quantify the exact change of the complexity between different quenches.
The expression for the complexity between time period $iT<t<(i+1)T$ can be written as \cite{ABHKM} :
$\mathcal{C}_s(t)=\frac{1}{2}\sqrt{\sum_{j=1}^{N}\left(\mathcal{A}_{sj}^2+\mathcal{B}_{sj}^2\right)}$, 
where ${A}_{sj}^2$ and ${B}_{sj}^2$ are functions of
$b_{j,i}(t_0), {\dot b}_{j,i}(t_0), b_{j,i+1}, {\dot b}_{j,i+1}$, and are given by
\begin{equation}\label{complexity_successive}
\begin{split}
\mathcal{A}_{sj}(t)=\ln\Bigg[\frac{|\Omega_{Tj}|}{|\Omega_{Rj}|}\Bigg]~,~
\mathcal{B}_{sj}(t)=\arctan\bigg[\frac{\text{Im}(\Omega_{Tj})\text{Re}(\Omega_{Rj})-\text{Im}(\Omega_{Rj})\text{Re}(\Omega_{Tj})}
{\text{Re}(\Omega_{Rj})\text{Re}(\Omega_{Tj})+\text{Im}(\Omega_{Rj})\text{Im}(\Omega_{Tj})}\bigg]~.
\end{split}
\end{equation}
Here, $\Omega_{Tj}$ is the complex frequency of the target Gaussian state,
of the form of $\Omega_{j}$ given in Eq. (\ref{groundstate}),
and $\Omega_{Rj}$ is the frequency of reference state: $\Omega_{Rj}=\Omega_{j}(t_0)$, where $iT<t<(i+1)T$
and $(i-1)T<t_0<iT$ is  a fixed value of time at $i$th quench.
The expressions for these functions
$\mathcal{A}_{sj}(t)$ and $\mathcal{B}_{sj}(t)$ can be written in terms of the auxiliary function
$b_{j,i}(t_0)$ and $b_{j,i+1}(t)$ and their derivative  as
\begin{equation}
\mathcal{A}_{sj}(iT<t<(i+1)T)=\ln\Bigg[\frac{b_{j,i}^2(t_0)\sqrt{b_{j,i+1}^2(t)\dot{b}_{j,i+1}^2(t)
+\lambda_{(i/f)j}}}{b_{j,i+1}^2(t)\sqrt{b_{j,i}^2(t_0)\dot{b}_{j,i}^2(t_0)+\lambda_{(f/i)j}}}\Bigg]~,
\end{equation}
and 
\begin{equation}
\mathcal{B}_{sj}(iT<t<(i+1)T)=\arctan\Bigg[\frac{b_{j,i}(t_0)\dot{b}_{j,i}(t_0)\sqrt{\lambda_{(f/i)j}}
-b_{j,i+1}(t)\dot{b}_{j,i+1}(t)\sqrt{\lambda_{(i/f)j}}}{\sqrt{\lambda_{(i/f)j}}
\sqrt{\lambda_{(f/i)j}}+b_{j,i}(t_0)\dot{b}_{j,i}(t_0)b_{j,i+1}(t)\dot{b}_{j,i+1}(t)}\Bigg]~.
\end{equation}
Note that here $b_{j,i}(t_0)$ and $\dot{b}_{j,i}(t_0)$ are constants, indicating the value of the auxiliary
function $b_{j,i}(t)$ and its derivative at a time $t_0$. 
Furthermore, during the time period $t_0<t<iT$, i.e. between time 
$t_0$ and that of the $i$th quench the expression for the complexity is given by a similar formula 
as that of above with 
with $b_{j,i+1}(t)$ and $\dot{b}_{j,i+1}(t)$ respectively replaced by the auxiliary function $b_{j,i}(t)$
and its derivative respectively. Thus the functions $\mathcal{A}_{sj}$ and $\mathcal{B}_{sj}$ in this case are given by
\begin{equation}
\mathcal{A}_{sj}(t_0<t<iT)=\ln\Bigg[\frac{b_{j,i}^2(t_0)\sqrt{b_{j,i}^2(t)\dot{b}_{j,i}^2(t)
+\lambda_{(i/f)j}}}{b_{j,i}^2(t)\sqrt{b_{j,i}^2(t_0)\dot{b}_{j,i}^2(t_0)+\lambda_{(i/f)j}}}\Bigg]~,
\end{equation}
and 
\begin{equation}
\mathcal{B}_{sj}(t_0<t<iT)=\arctan\Bigg[\frac{b_{j,i}(t_0)\dot{b}_{j,i}(t_0)\sqrt{\lambda_{(i/f)j}}
-b_{j,i}(t)\dot{b}_{j,i}(t)\sqrt{\lambda_{(i/f)j}}}{\sqrt{\lambda_{(i/f)j}}
\sqrt{\lambda_{(i/f)j}}+b_{j,i}(t_0)\dot{b}_{j,i}(t_0)b_{j,i}(t)\dot{b}_{j,i}(t)}\Bigg]~.
\end{equation}


\begin{figure}[h!]
\begin{minipage}[b]{0.45\linewidth}
\centering
\includegraphics[width=0.95\textwidth]{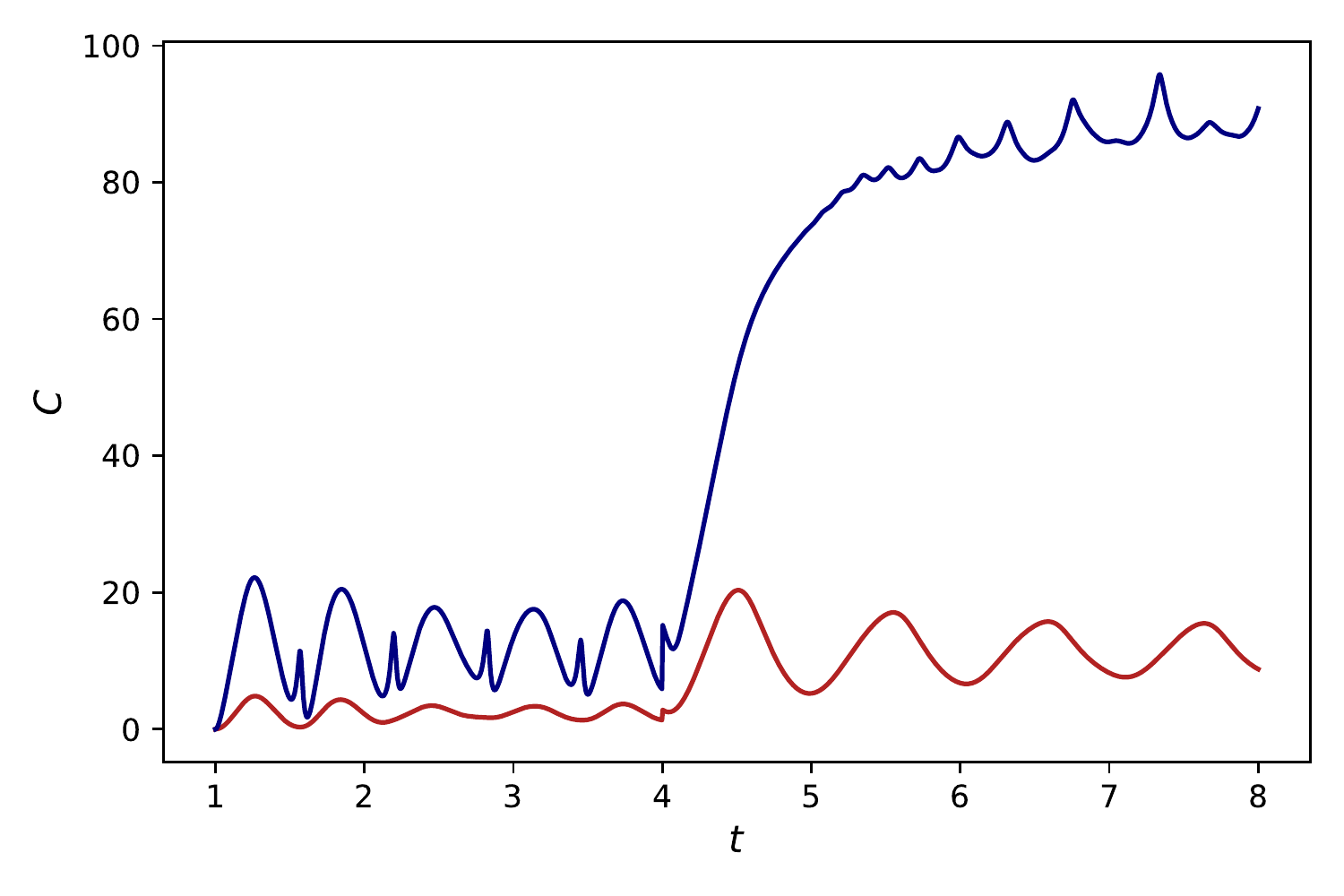}
\caption{Time evolution of complexity between first and second quench,
with $N=100$ and $k_f=k_i=4,T=4$. Blue: $\omega_f=5$ and $\omega_i=0.3$,
red:  $\omega_i=3$ and $\omega_f=5$.}
\label{fig:successive_12}
\end{minipage}
\hspace{0.2cm}
\begin{minipage}[b]{0.45\linewidth}
\centering
\includegraphics[width=0.95\textwidth]{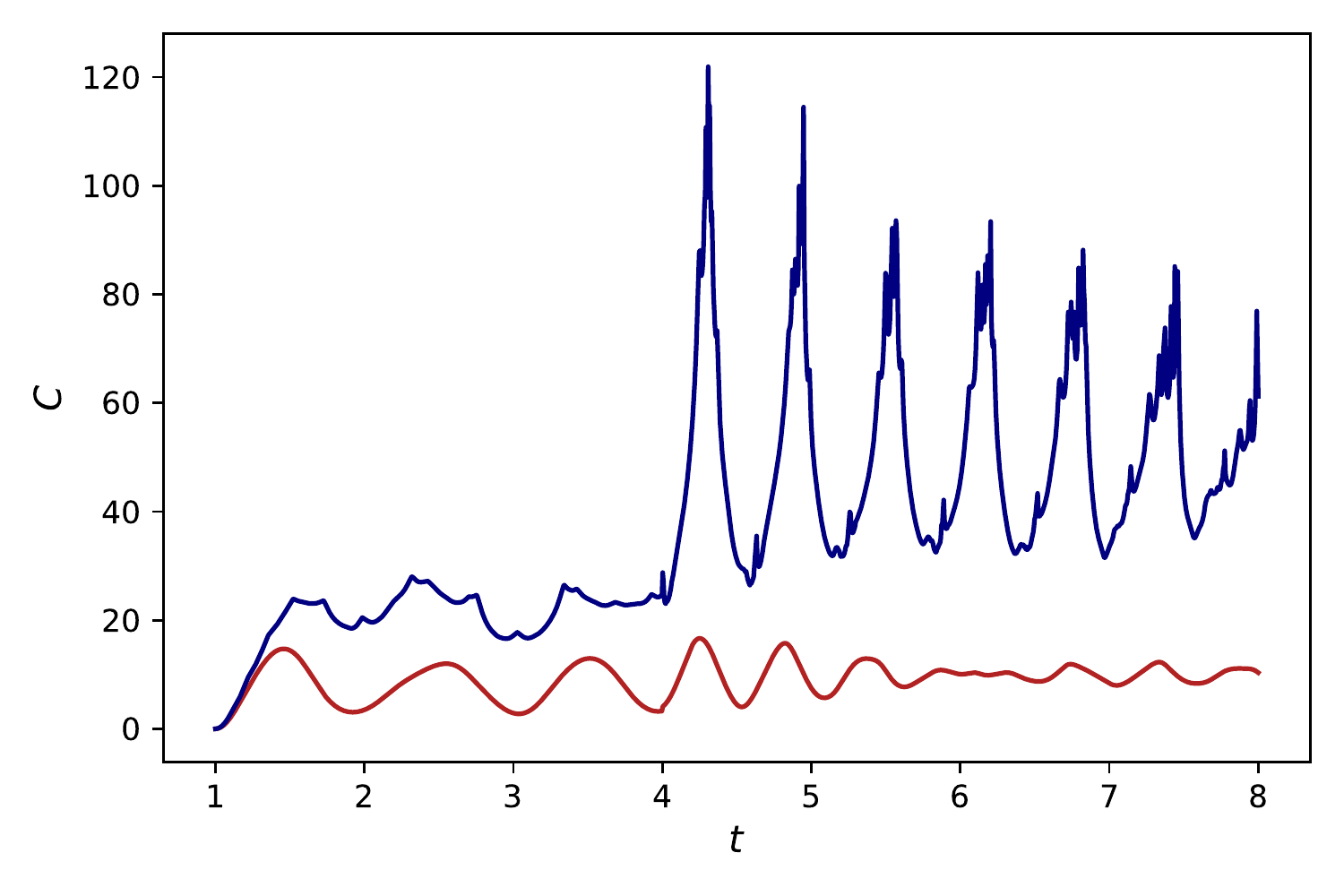}
\caption{Time evolution of complexity between  second and third quench,
with $N=100$ and $k_f=k_i=4,T=4$. Blue: $\omega_f=5$ and $\omega_i=0.3$,
red:  $\omega_i=3$ and $\omega_f=5$. }
\label{fig:successive_23}
\end{minipage}
\end{figure}

Time evolution of complexity between the first and second quenches is shown in Fig. \ref{fig:successive_12},
and between the second and third quenches is shown in
Fig. \ref{fig:successive_23}, with Fig. \ref{fig:successive_23}
being shifted appropriately in time.
In Fig. \ref{fig:successive_12}, the reference state is chosen to be a 
state at fixed time $t_0=1$
after the first quench and the target state is a state at an arbitrary time 
$t$ between $t_0<t<2T$ where $T=4$. While in Fig. \ref{fig:successive_23},
the reference state is taken at $t_0=5$, just after the second quench
and the target state is a state is between $t_0=5<t<3T$. In both the cases 
the blue curve is with $\omega_i=0.3,\omega_f=5$ and the red one with 
$\omega_i=3,\omega_f=5$.

The first point to note is that in both the cases, just after the
$i+1$ th quench, there is a sharp rise in the complexity, which was absent 
when the complexity was calculated from the fixed state before the quench.
When the difference between the target state and reference state frequencies are higher, the 
growth of the complexity after the quench is sharper.
This growth can be characterised by expanding the complexity in Eq. 
(\ref{complexity_successive}) in powers of $t$ as in Eq.  (\ref{muliple_expansion})
and finding out the coefficients of various order contributions.

As we have mentioned before, when the reference state is a fixed at $t=0$, for a time
period just  after any quench, say with a quench number $i$, the complexity has value 
lower than its value at $t<(i-1)T$.  For example, we see from Fig. \ref{fig:Complexity_3quenchs} 
that after the second quench (where $t=8$) the complexity increases and then takes a value 
lower than its value at $t=8$. 
However, unlike in Fig. \ref{fig:Complexity_3quenchs}, here, for the complexity
between states in successive quenches, the complexity
never comes below its  pre-quench value. Thus  when the reference
state is a state in the $i$th quench and the target state is in the $(i+1)$th quench, during 
the evolution after the $(i+1)$th quench, the target state never comes closer to the reference
state than the state during $i$th quench. Thus after the quench complexity sharply increases.
This is in sharp contrast with the complexity evolution the  with reference state before the quench.

\begin{figure}[h!]
\begin{minipage}[b]{0.45\linewidth}
\centering
\includegraphics[width=0.95\textwidth]{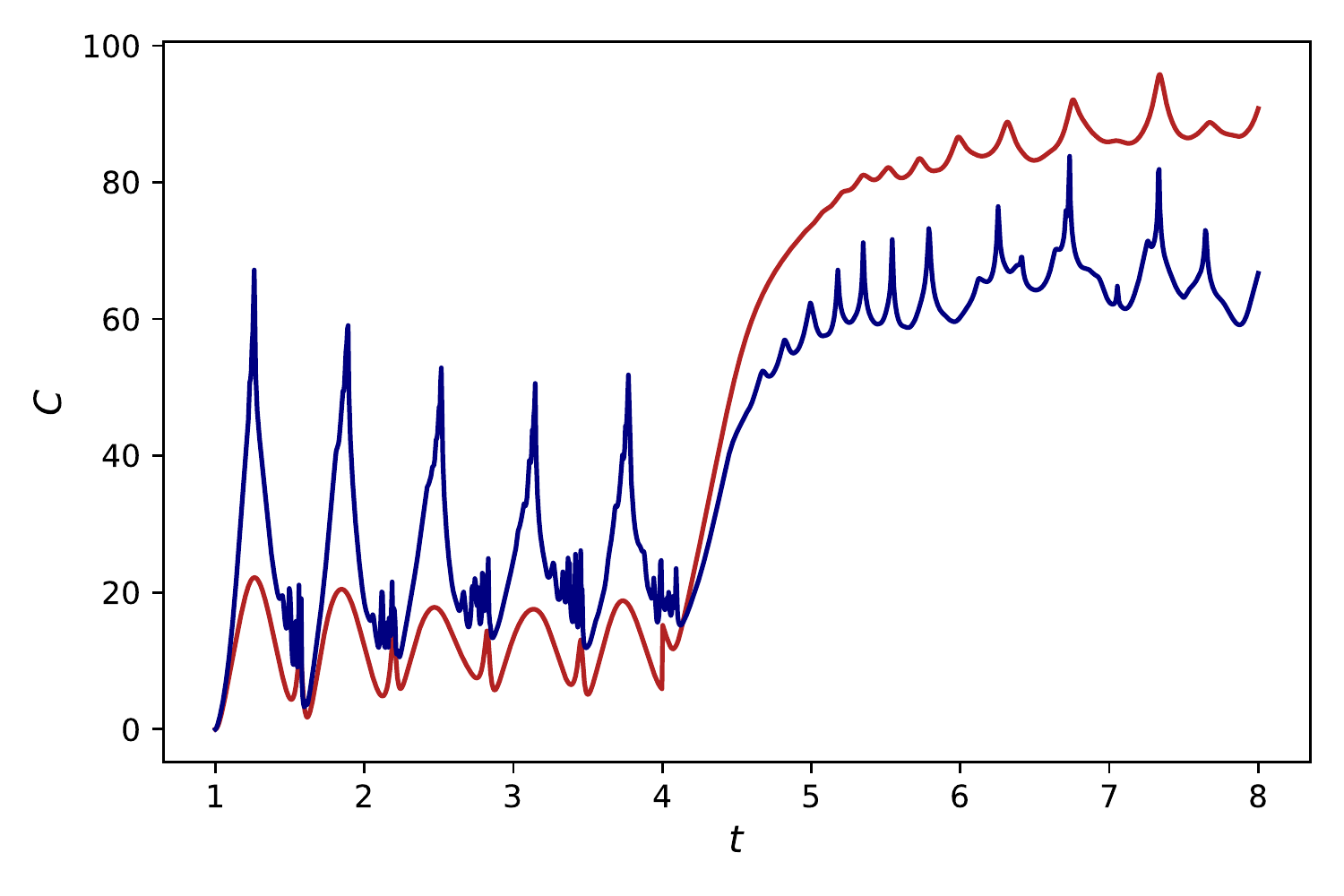}
\caption{Comparison  of complexities between first and second quench (red curve),
with that of third and fourth quench (blue), with $N=100$, $k_f=k_i=4,T=4$,
$\omega_f=5$ and $\omega_i=0.3$. }
\label{fig:successive__comparison}
\end{minipage}
\end{figure}

Another interesting phenomena can be observed  by comparing the complexities between two different successive
quenches. In Fig. \ref{fig:successive__comparison}, we have compared the complexities between 
the first and the second quench (red curve) to that of between the third and the fourth quench (blue curve). 
Here, the time scale of the blue curve has been shifted to make a comparison.  
Comparing these two curves, we see that before the $i+1$th quench ($i=1$ for red and $i=3$ for blue),
the complexity with higher value of $i$ is always higher. However just after the $i+1$th 
quench is performed, the complexity having lower value of $i$ crosses the complexity with higher $i$
and always stays higher up to their respective ranges of time. 
In terms of points on the space of unitary operators, we can interpret this phenomena as follows.
Let us take the states at two different $t_0$s on two $i$th quenches 
to be two fixed reference states on the unitary manifold and compute 
the distances of two time evolved states before the respective $(i+2)$th quenches
from these two reference states.  
Before the $(i+1)$th quenches are performed the target state which is located at later times 
is always at a further distance  from its reference state than the one at earlier
time. But ones the $(i+1)$th quench is performed this distance changes rapidly so that  the 
target state  at higher time comes much closer to its reference state than the one at lower time. 
Thus we can conclude that the effect a quench at an earlier  time is to `push away' the time evolved 
state on the space of unitary manifold compared with a target state at a later time. This phenomena 
can not be observed when a single quench is performed on the harmonic chain.

We also notice that this `crossover' of the complexities of successive quenches
can be most clearly observed when the  frequency $\omega_{f}$ is sufficiently higher  
than the initial frequency $\omega_i$. For these two frequencies having a value much closer 
to each other, the behaviour of complexity evolution is more intricate and 
it will be interesting to quantify this further. Furthermore, it will be interesting to 
see it this phenomena can be observed in complexity between successive quenches when
the covariance matrix is used to calculate the complexity instead of the wavefunction 
method that we have used. This might be helpful to distinguish these two methods further.

\section{Conclusions}

We have used the Lewis-Risenfeld  method to obtain the time evolution of
the circuit complexity in a time-dependent harmonic chain under sudden single or multiple quenches
of the frequency and coupling strength. Under a single quench, much like the entanglement
entropy, the complexity shows quasi-revivals due to the present of dynamically generated time scales,
whose number in turn depends on the total number of particles in the chain. When the quench is critical,
the complexity grows at large times due to the presence of a zero mode, however it does not show any kind of
decoherence in time, even for large post-quench frequency.

On the other hand, when multiple quenches are applied to the chain, though
the complexity increases sharply after each quench,
however even at late times it does not acquire any steady state value,
which is  observed in the OTOC  - indicating scrambling of information in the chain. 
For multiple quenches, several new features in the complexity are observed,
which can not seen in the single quench scenario.
Firstly, even after two successive quenches, when the frequency return to
its initial value, there is a lower limit in the value the complexity can have, which 
cannot be made to approach zero even by increasing the frequency sufficiently. 
Secondly, by applying a large number of successive quenches to the chain, the
complexity of the time evolved state can be increased to a high value, which
is not possible by applying a single quench. Finally, the complexity calculated 
between successive quenched states, having sufficient difference in frequency
shows an interesting behaviour, namely,
a quench  performed at an earlier  time (say with a quench number $i$)
`pushes away' the time evolved target state further on the space of unitaries 
compared to a quench performed at a later time. Thus we observe a crossover of the 
two complexities just after the $(i+1)$th quench.

In this paper, although our analysis is for a sudden quench, the formalism can
be applied to more general time-dependent frequencies or interaction profiles, 
to study the time-evolution of complexity. Furthermore, it will be interesting
see weather the special features of the complexity after multiple quenches
listed above are shared by other systems as well.
	

\end{document}